\documentstyle[psfig,proceedings]{crckapb}

\begin{opening}
\title{GRAVITATIONAL LENSES AND THE HUBBLE CONSTANT:\protect\\
       PRESENT AND FUTURE}
\author{L.J. GOICOECHEA}
\institute{Departamento de Fisica Moderna, Universidad de Cantabria\\
           Avda. de los Castros, E-39005 Santander, Cantabria (Spain)\\
	   E-mail: goicol@besaya.unican.es}
\author{E. MEDIAVILLA}
\author{A. OSCOZ}
\author{M. SERRA}
\author{J. BUITRAGO}
\institute{Instituto de Astrofisica de Canarias (IAC)\\
           Via Lactea, E-38200 La Laguna, Tenerife (Spain)\\
	   E-mail: emg@iac.es,aoscoz@iac.es,mserra@iac.es,jgb@iac.es}
\end{opening}

\runningtitle{GRAVITATIONAL LENSES AND $H_0$}

\begin{document}

\section{Introduction}

For a multiple QSO, the propagation time from
the source to the observer varies from the image $i$ to the image $j$, and this
difference ($\Delta\tau_{ij}$) can be measured when the source is variable.
In general, assuming a flat universe without cosmological constant, the 
parameter $\Delta\tau_{ij}$ x $H_0$ ($H_0$ is the Hubble constant) depends on 
the redshifts of the lens and
the source, as well as the positions of the individual images and the source
($\vec \theta_{i},\vec \theta_{j},\vec \beta$), and the scaled surface 
potential $\Psi_{\alpha}$ at $\vec \theta_{i}$ and $\vec \theta_{j}$ (see,
e.g., Blandford and Kundic 1996, Williams and Schechter 1997). The 
observations of multiple images of the same source are used to infer 
$\vec \beta$ and the adjustable parameters 
$\alpha \equiv (\alpha_{1},...,\alpha_{p})$ that appear in the picture of the
deflector, i.e., a lens model. From the lens model corresponding to the lens
picture, $\Delta\tau_{ij}$, $\vec \theta_{i}$, $\vec \theta_{j}$ and the
redshifts, one obtains $H_0$. 

A golden system (which is suitable for determining $H_0$) must be
a multiple QSO verifying some properties. The deflector cannot be
dark or very faint (non-observable), and in order to model the 
gravitational potential, the lensing structure must be simple (e.g., an 
isolated galaxy). To infer $\alpha$ and
$\vec \beta$, a relatively large number of constraints is also desirable. 
On the other hand, for obtaining $\Delta\tau_{ij}$, some typical scales of 
intrinsic
variability of the source should be less than $\Delta\tau_{ij}$, and
the absence of strong short timescale microlensing is need.
The microlensing events mask the intrinsic variability.

The best studied gravitational lens is the {\it double} QSO Q0957+561A,B. As
a result of recent advances in observations and modelling, a ten percent
measurement of $H_0$ is attainable. We discuss in detail this system (Sect. 2)
and review the perspectives in a near future (Sect. 3). 

\section{Twin QSO (0957+561$A,B$)}

For Q0957+561$A,B$, at optical wavelengths,
there are epochs of a rapid intrinsic variability (a basic condition for
measuring the time delay accurately) and epochs of calmness. They also appears
evidences in favour of an important short timescale microlensing (Schild 1996),
which masks the intrinsic variability. So, due to the double behaviour 
activity/calmness, the microlensing events, the gaps in the monitoring (three
months per year), etc., it is difficult an accurate time delay determination
based on a large dataset containing several years of observations.

The rough estimation $\Delta\tau_{BA} \approx T$ = 420 days (Pelt {\it et al.}
1996; see also the paper by Pijpers in this volume) can be used for making nice
datasets. A nice dataset contains an {\it active} light curve $A$ during a 
period
[$t_{i},t_{f}$] and the light curve $B$ in the interval [$t_{i}+T,t_{f}+T$]. The
microlensing should be weak and, on the other hand, the absence of gaps (each
summer) is need. Very recently, using image $A$ data in the band $r$ (1994
Dec.-1995 May, which were kindly provided us by T. Kundic), and our image $B$
data in the band $R$ (1996 Feb.-July), we concluded that 
$\Delta\tau_{BA}$ = 424$\pm$3 days (Oscoz {\it et al.} 1997). The $A$ component
was active (Kundic {\it et al.} 1995), and
also, we have not found evidences of strong microlensing in the $r-R$ 
comparison (Oscoz {\it et al.} 1997, Goicoechea {\it et al.} 1998). A similar
result (417$\pm$3 days; 2$\sigma$ confidence level) has been derived by Kundic 
{\it et al.} (1997). They use light curves in
the $g$ and $r$ bands. However, a reanalysis based on discrete correlation
functions and two nice datasets, $A_{r}+B_{r}$ and $A_{g}+B_{g}$ (we basically
exclude the photometric monitoring of image $B$ in the interval 1995 Dec.-1996
Jan.), shows that 424 days is the most probable delay. For every fixed
value $\theta$ (days), we construct the function 
\begin{equation}
\delta^2(\theta) = (1/N)\sum_{i=1}^NS_{i}[DCC(\tau_{i}) -
DAC(\tau_{i}-\theta)]^2 ,
\end{equation}
where $DCC$ is the discrete $A-B$ cross-correlation function, $DAC$ is the
discrete $A-A$ autocorrelation function and $S_{i}$ = 1 only when both $DCC$
and $DAC$ are defined at $\tau_{i}$ and $\tau_{i}-\theta$, respectively, and 
$S_{i}$ = 0 otherwise. Then, we search for a minimum $\theta_{0}$, such that 
$\theta_{0} = \Delta\tau_{BA}$. In the two ($g-g$ and $r-r$) comparisons, 
it is inferred $\Delta\tau_{BA}$ = 424 days (see Fig. 1).

\begin{figure}
%\vspace{7cm}  % amount of vertical space needed
\psfig{file=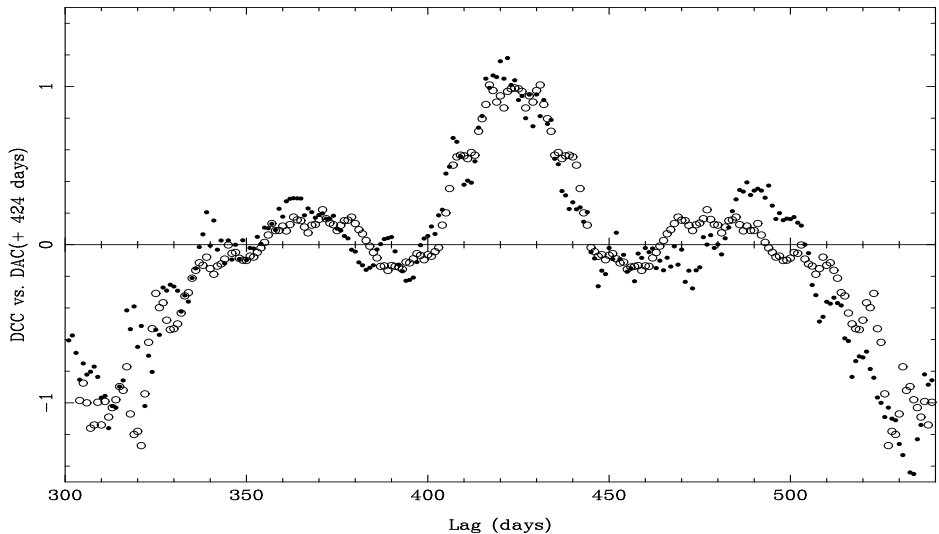,bbllx=70pt,bblly=30pt,bburx=570pt,bbury=700pt,%
height=7cm,width=12cm,angle=270}
\caption{$DCC$ (filled circles) vs. $DAC$ (shifted by 424 days; open circles)
in the $g-g$ comparison. The agreement is excellent (for details, see main 
text).}
\end{figure}
The main deflector is a cD galaxy at the center of a cluster. 
A good picture of the galaxy could be
a King-type surface density profile with velocity dispersion $\sigma_v$ and
core angular radius $\theta_c$, plus a point-mass ($M_{bh}$) at the centre of
the King profile (Falco {\it et al.} 1991). Moreover, the gravitational effect 
of the cluster can be represented by means of a
quadratic lens with a convergence $\kappa$ and a shear $\gamma$ with position
angle $\phi$. Grogin and Narayan (1996) used 
$\alpha \equiv (\sigma_{v},\theta_{c},M_{bh},\gamma,\phi)$,
$\vec \beta_{1} \equiv (\beta_{1x},\beta_{1y})$,
$\vec \beta_{5} \equiv (\beta_{5x},\beta_{5y})$ as free parameters (Garrett
{\it et al.} 1994, have fitted the $A$ and $B$ radio images with six Gaussian
components, being $A_1$ and $B_1$ the respective core components and
$A_5$-$B_5$ another jet components), since, in spite of the large number of
constraints, there is a degenerancy in the convergence due the cluster. The
whole lens model must include the nine parameters inferred from measurements of
the lensed images as well as an estimation of $\kappa$ derived from a direct
measurement of the mass either in the cD galaxy or the cluster. A measurement
of $\sigma_{v}(light)$, the 1D velocity dispersion of the luminous stars in the
galaxy, allows to eliminate the cluster degenerancy. 

Falco {\it et al.} (1997) derived $\sigma_{v}(light)$ = 279$\pm$13 km/s,
and so, with the whole lens model and the time delay above
mentioned, one obtains $H_0$ = 66$^{+15}_{-14}$ km/s/Mpc (2$\sigma$). This
estimate is 10$\%$ accurate at 1$\sigma$ 
(Oscoz {\it et al.} 1997). We however remark that
a new lens model satisfying the constraints from radio mapping as well as new
optical constraints deduced from {\it HST} observations is
now in progress (Bernstein {\it et al.} 1997). Also, the 
approximation  $\sigma_{v} = \sigma_{v}(light)$ must be reconsidered
(Mediavilla {\it et al.} 1998).

\section{Perspectives in a near future}

The combined effort by several groups of astronomers will lead to the detailed
analysis of a large number of gravitational mirages. For each system, if the
deflectors are not dark, we must be able to obtain a
promising set of constraints and some information on suitable pictures of the
global lens (primary and secondary deflectors). Through a good picture and the
constraints, one can infer a good lens model. Moreover,
the system must be extensively monitored to get the light curves of the
different images and deduce at least one time delay. This work could be
difficult due to probable microlensing events, a weak source variability, an
unsuitable sampling of the image light curves, etc. The technique for
determining a time delay plays also a role. Even in the {\it old} Twin QSO
there is an uncertainty (due to the methodology) of about one week, which is
irrelevant in order to obtain $H_0$ from this system, but it could be dramatic
in another multiply imaged QSO (e.g., the Triple QSO
1115+080$A_{1}-A_{2},B,C$).

Nowadays, from two individual systems (Twin QSO and Triple QSO), it is
inferred a mean value of $H_0 \approx$ 60 km/s/Mpc, in good agreement with
other methods (see the paper by C. Frenk in this volume). However, new
measurements of $H_0$ from gravitational lenses may surprise us. For example,
when one measures $H_0$ via gravitational lensing, the influence of the
large-scale structure is not taken into account. The effect introduced in the
measurement of $H_0$ caused by large clusters and/or large voids, is a very
interesting topic, which will be soon studied by our group.

\section{References}

Bernstein {\it et al.} (1997) {\it ApJ} {\bf 483}, L79\\ 
Blandford, R.D. and Kundic, T. (1996) {\it preprint (astro-ph/9611229)}\\
Falco {\it et al.} (1991) {\it ApJ} {\bf 372}, 364\\  
Falco {\it et al.} (1997) {\it ApJ} {\bf 484}, 70\\ 
Garrett {\it et al.} (1994) {\it MNRAS} {\bf 270}, 457\\  
Goicoechea {\it et al.} (1998) {\it ApJ} {\bf 492}, 74\\  
Grogin, N.A. and Narayan, R. (1996) {\it ApJ} {\bf 464}, 92; {\bf 473}, 570\\  
Kundic {\it et al.} (1995) {\it ApJ} {\bf 455}, L5\\  
Kundic {\it et al.} (1997) {\it ApJ} {\bf 482}, 75\\  
Mediavilla {\it et al.} (1998) {\it in preparation}\\  
Oscoz {\it et al.} (1997) {\it ApJ} {\bf 479}, L89\\  
Pelt {\it et al.} (1996) {\it A$\&$A} {\bf 305}, 97\\ 
Schild, R. (1996) {\it ApJ} {\bf 464}, 125\\  
Williams, L.L.R. and Schechter, P.L. (1997) {\it preprint (astro-ph/9709059)}

\end{document}